    \documentclass[twocolumn,prl,showpacs,english,aps]{revtex4-1}
    \usepackage{babel,rotating,dcolumn}
    \usepackage{colordvi,graphicx,color,amsbsy,amsmath,bm,amssymb}
    \usepackage{comment}
    \usepackage{mathrsfs}

    \usepackage{colordvi,color} 
    \usepackage{xcolor}
    \usepackage{hyperref}

\begin{document}
    
    \preprint{APS/123-QED}

    \title{Non-diffusive transport  of Inertial Heavy Impurities in Drift-Wave Turbulence}

    \author{Zetao Lin}
    \affiliation{Aix-Marseille Universit\'e, CNRS, I2M, UMR 7373, 13331 Marseille, France}
    \author{Benjamin Kadoch}%
    \affiliation{Aix-Marseille Universit\'e, CNRS, IUSTI, UMR 7343, 13453 Marseille, France
    }
    \author{Sadruddin Benkadda}
    \affiliation{Aix-Marseille Universit\'e, CNRS, PIIM, UMR 7345, 13397 Marseille, France}%
    \author{Kai Schneider}
  
    \affiliation{Aix-Marseille Universit\'e,  CNRS,  I2M , UMR 7373, 13331 Marseille, France}
    \date{\today}
    \begin{abstract}
    We investigate the transport behavior of tungsten impurities with finite inertia in drift-wave turbulence using the Hasegawa--Wakatani model. Unlike previous tracer-based models, our simulations reveal a transition to non-diffusive dynamics for a range of charge states. This novel mechanism offers a  turbulence-driven
    route to core impurity accumulation. This finding underscores the nontrivial role of particle inertia in impurity dynamics and has strong implications for impurity control in future fusion devices such as ITER.

    \end{abstract}
    
    \maketitle
    Impurities play complex roles in fusion plasmas, influencing both device performance and operational stability. Controlled impurity injection is anticipated to be essential for divertor radiative cooling and disruption mitigation in next-generation fusion devices such as ITER and DEMO. However, unintended accumulation of impurities in the plasma core increases radiation losses and fuel dilution, thereby reducing energy confinement and fusion performance \cite{Pitts2013, Hollmann2015, Angioni2021, Kallenbach2013}. Consequently, developing a precise understanding of impurity transport dynamics becomes fundamental for enhancing operational capabilities in magnetic confinement fusion devices. Within tokamaks and similar fusion systems, transport phenomena typically exhibit anomalous characteristics rather than classical collision-based processes \cite{Hidalgo1995, Naulin2007,horton1999drift}. Anomalous transport in magnetically confined fusion plasmas often deviates from simple diffusive models, particularly in the near-marginal conditions where profiles hover near instability thresholds. A  paradigm for understanding these dynamics is  Self-Organized Criticality (SOC), which describes transport as a series of intermittent, avalanche-like events that exhibit non-local correlations and long-term memory. The evidence for this SOC-like behavior in both simulations and experiments is extensively reviewed \cite{Sanchez2015}. Within this SOC framework,  the  non-diffusive transport in plasma turbulence is often related to  near-marginality or with the presence of strong, sheared zonal flows \cite{Sanchez2008}. This Letter identifies a new  mechanism for non-diffusive transport. By analyzing impurity trajectories in a supermarginal Hasegawa--Wakatani system, we provide quantitative evidence that inertia induces a transition from diffusive to  non-diffusive transport. providing a previously missing mechanism for rapid core accumulation of tungsten in ITER-like plasmas.
    
In tokamak edge regions, drift-wave turbulence is primarily responsible for the turbulent transport of particles and heat~\cite{horton1999drift}. The Hasegawa--Wakatani (HW) model is a valuable tool for investigating  drift-wave turbulence, offering both theoretical robustness and computational efficiency~\cite{Hasegawa1983, camargo1995, Lin2024}. However, a gap exists in current plasma physics research regarding the behavior of impurities with finite inertia when modeled within the HW framework.  
    
    Prior research has examined passive tracer dynamics in plasma turbulence, but has largely neglected inertial effects~\cite{Futatani2008, Bos2010, Sanchez2006}.  While the importance of particle inertia in causing complex transport behaviors is recognized in classical fluid dynamics~\cite{Gustavsson2016}, its consequences in a plasma turbulence context remain largely unexplored. The zero-inertia approximation becomes problematic when considering next-generation fusion devices such as ITER, where tungsten and other heavy impurities will play a significant role. To address this knowledge gap, we present in this letter a comprehensive numerical study of impurity transport in HW drift-wave turbulence that explicitly accounts for particle inertia.  Our approach systematically characterizes transport behavior across a series of Stokes numbers through one-way coupling between the turbulent plasma flow and impurity particles. We employ Lagrangian tracking of individual impurity particles through a   drag formulation. We propose that inertial effects cause heavy impurities to exhibit a delayed response to the surrounding plasma flow, creating a characteristic ``lag'' phenomenon. 
    
    To simulate the turbulent environment, we generate plasma fields by numerically integrating the two-dimensional HW equations. These equations capture the essential physics of electrostatic potential, $\phi$ and density fluctuations, $n$, in the tokamak edge region~\cite{Hasegawa1983,Kadoch2010}:
    \begin{align}
    &(\frac{\partial}{\partial t}-\mu_\nu \nabla^2) \nabla^2 \phi=[\nabla^2 \phi, \phi]+c(\phi-n) \\ 
    &(\frac{\partial}{\partial t}-\mu_D \nabla^2) n=[n, \phi]-\Gamma \frac{\partial \phi}{\partial y}+c(\phi-n)
    \end{align}
    where $\mu_D$ is the cross-field diffusion coefficient and $\mu_\nu$ is the kinematic viscosity. The term $\Gamma$, defined as $\Gamma \equiv-\partial_x \ln \left(n_0\right)$, measures the plasma density gradient. The Poisson bracket is defined as: $[A, B]=\frac{\partial A}{\partial x} \frac{\partial B}{\partial y}-\frac{\partial A}{\partial y} \frac{\partial B}{\partial x}$. In these equations, the electrostatic potential $\phi$ is the stream function for the $\boldsymbol{E} \times \boldsymbol{B}$ velocity, represented by $\boldsymbol{u}=\nabla^{\perp} \phi$, thus we have $u_x=-\frac{\partial \phi}{\partial y}$ and $u_y=\frac{\partial \phi}{\partial x}$. The vorticity is given by $\omega=\nabla^2 \phi$. The variables are normalized as in~\cite{Bos2010, Kadoch2010}.
    The adiabaticity parameter $c$ measures the parallel electron response and is defined as $
    c=(T_e k_{\|}^2)/(\mathrm{e}^2 n_0 \eta \omega_{c i})$ where $\eta$ is the electron resistivity and $k_{\|}$is the effective parallel wavenumber. We performed DNS computations using a fully dealiased pseudo-spectral method  at resolution $N_x \times N_y = 1024^2$ grid points in a double periodic domain of size length 64 over 400 eddy turnover times. For details, we refer to~\cite{Lin2025a}.

    The dynamics of heavy impurity particles in plasma is governed by two primary forces: the Lorentz force arising from electromagnetic fields and a collisional drag force resulting from interactions with background plasma ions. While momentum exchange with plasma ions occurs through the electrostatic potential surrounding the impurity, the contribution from electrons can be neglected due to their substantially smaller mass. The equation of motion for these heavy impurity particles can be expressed as:
    \begin{equation}
    m_p \frac{d \mathbf{v}_p}{d t} = \mathbf{F}_{\text{drag}} + \mathbf{F}_{\text{Lorentz}},
    \end{equation}
    with $m_p$ and $\mathbf{v}_p$ respectively being the particle mass and the velocity, which expands to:
    \begin{equation}
    \frac{d \mathbf{v}_p}{d t} = \frac{\mathbf{u}_p - \mathbf{v}_p}{\tau_p} + \frac{Z e}{m_p}(\mathbf{E} + \mathbf{v}_p \times \mathbf{B}).
    \end{equation}
    The characteristic relaxation time $\tau_p$, which we derived in our earlier research \cite{Lin2025b}, is given by:
    \begin{equation}
    \tau_p = \frac{m_p}{\sqrt{m_i}} \frac{6\sqrt{2}\pi^{3/2}\epsilon_0^2}{n_i Z^2 e^4} \frac{T_i^{3/2}}{\ln\Lambda}\, ,
    \end{equation}
    where $n_i$ is the plasma ion density; $T_i$ is the ion temperature; $\ln \Lambda$ is  Coulomb logarithm. When normalized in accordance with the HW framework, the particle motion equations take the form:
    \begin{equation}
    \frac{d \boldsymbol{v}_p}{d t} = \frac{\boldsymbol{u}_p - \boldsymbol{v}_p}{\tau_p} + \alpha(-\nabla \phi(\boldsymbol{x}_p) + \boldsymbol{v}_p \times \boldsymbol{b}),
    \label{eq: dv dt}
    \end{equation}
    with the parameter $\alpha = Zm_i/m_p$.
    
    The spatial evolution of impurity particles is determined by:
    \begin{equation}
    \frac{d \boldsymbol{x}_p}{d t} = \boldsymbol{v}_p.
    \label{eq: dx dt}
    \end{equation}
    Equations (\ref{eq: dv dt}) and (\ref{eq: dx dt})  provide a complete description of impurity trajectories in the turbulent plasma environment.
    
    We focus on tungsten impurities originating from plasma-facing components in fusion devices such as ASDEX Upgrade \cite{Krieger1999}, EAST \cite{Yao2015}, and ITER \cite{Pitts2009}. Various tungsten charge states ($Z=3,10,20,60$) represent low to highly charged ions, and corresponding relaxation times and Stokes numbers ($St=\tau_p/\tau_\eta$) are systematically investigated. The Stokes number is a dimensionless parameter that characterizes the behavior of particles suspended in a fluid flow, representing the ratio of the particle's relaxation time $\tau_p$ to a characteristic time scale of the flow $\tau_{\eta}$. When $St \ll 1$, particles closely follow fluid streamlines, while particles with $St \gg 1$ respond minimally to flow fluctuations due to their significant inertia. Here, $\tau_\eta=0.35$ is determined by the turbulent vorticity field. The parameters to calculate $\tau_p$ include $n_i = 1 \times 10^{19},\mathrm{m}^{-3}$,  $T_i = 0.1 \mathrm{eV}$, $\ln \Lambda = 10$.  The normalized relaxation times and $\alpha$ for different values of $Z$ for tungsten ions are presented in Table~\ref{tab:taup_alpha}.  

\begin{figure}
      \includegraphics[width=0.8\columnwidth]{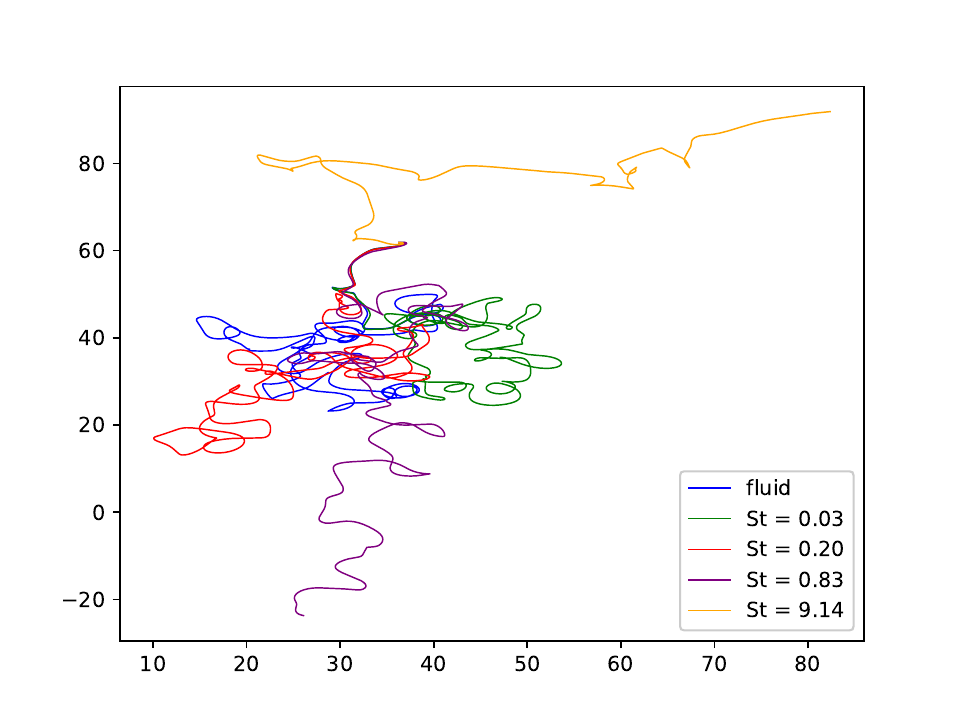}
      \caption{\label{fig:trajectory}Trajectories for fluid tracers ($St = 0$) and impurity particles with different Stokes numbers are shown. The periodic domain $[0, 64] \times [0, 64]$ has been extended to display continuous trajectories. The common initial position is marked with a green cross.}
    \end{figure}

    \begin{table}[t]
    \caption{\label{tab:taup_alpha}Normalized relaxation time $\tau_p$, $\alpha$, and Stokes number for tungsten ions.}
    \begin{ruledtabular}
    \begin{tabular}{ccccc}
    $Z$ & 3 & 10 & 20 & 60 \\
    \hline
    $\tau_p$ & 3.20 & 0.29 & 0.07 & 0.01 \\
    $\alpha$ & 0.03 & 0.11 & 0.22 & 0.66 \\
    $St$ & 9.14 & 0.83 & 0.20 & 0.03 \\
    \end{tabular}
    \end{ruledtabular}
    \end{table}
 \begin{table}[t]
\caption{\label{tab:exponents}The Mean Square Displacement (MSD) exponent $\mu$, the trapping time exponent $\gamma$, and the Hurst exponents $H$ for the velocity.}
\begin{ruledtabular}
\begin{tabular}{ccccccc}
$St$ & & 0 (fluid) & 0.03 & 0.20 & 0.83 & 9.14 \\
\hline
$\mu$ & & 0.97 & 0.97  & 1.10 & 1.32 & 1.21 \\
$\gamma$ & & -1.65 & -1.93 & -2.63 & -3.37 & -3.19 \\
$H$ & & 0.52 & 0.52 & 0.55 & 0.62 & 0.81 \\
\end{tabular}
\end{ruledtabular}
\end{table}
    Figure~\ref{fig:trajectory} illustrates representative trajectories of both fluid tracers ($St=0$) and impurity particles with varying Stokes numbers, each originating from an identical initial position (indicated by a green cross). To enhance visual clarity, we extended the periodic computational domain $[0,64]\times[0,64]$ to reveal uninterrupted particle pathways. This same periodic extension technique was employed in our statistical analyses. Fluid tracers ($St=0$) and particles with low Stokes number ($St=0.03$) exhibit similar random motion characteristic of advection by the turbulent flow. As the Stokes number increases, significant changes in particle behavior emerge. At $St=0.83$, particles are intermittently trapped in flow regions before abruptly jumping to new locations.  For particles with high Stokes number ($St=9.14$), inertia dominates their dynamics; they are less influenced by the fluid flow and tend to maintain their momentum, resulting in straighter trajectory segments between occasional flow interactions.

\begin{figure}[h!]
      \includegraphics[width=\columnwidth]{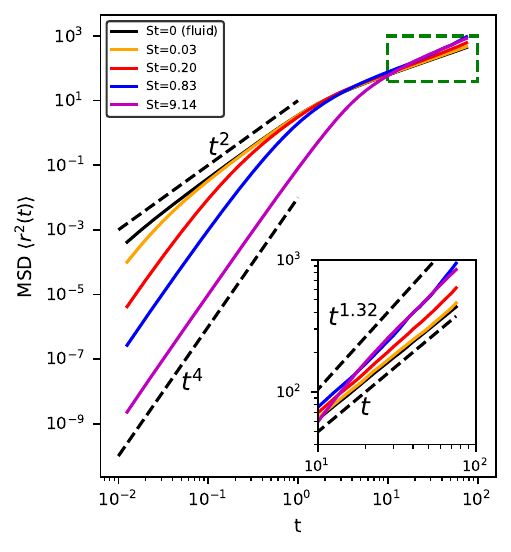}
      \caption{\label{fig:dispersion}Mean square displacement versus time for particles with different Stokes numbers. The inset highlights the long-time regime. Fluid tracers ($St=0$) show ballistic behavior at short times followed by normal diffusion, while particles with higher Stokes numbers exhibit hyperballistic motion at short times and superdiffusive behavior at long times.}
    \end{figure}

    The mean square displacement (MSD) $\left\langle r^2(t)\right\rangle$ as a function of time measures how far, on average, particles move from their initial position over time.  It is defined as $\left\langle r^2(t)\right\rangle = \left\langle |{\bf r}(t) - {\bf r}(0)|^2 \right\rangle$, where ${\bf r}(t)$ is the position of a particle at time $t$ and the brackets denote an ensemble average over particles.  Figure~\ref{fig:dispersion} illustrates how the  MSD  evolves over time for particles across various Stokes numbers, highlighting fundamentally different transport mechanisms. Fluid tracers ($St=0$)  initially following ballistic motion ($\left\langle r^2(t)\right\rangle$  $\propto t^2$)  where  displacement follows ${\bf r}(t) - {\bf r}(0) = {\bf v}_0 t$ at short timescales up to $t =1$. Then they transit to normal diffusion ($\left\langle r^2(t)\right\rangle$  $\propto t$) when $t > 10$, in agreement with the results reported in~\cite{Bos2010}. Particles with very low inertia ($St=0.03$) demonstrate transport properties virtually indistinguishable from pure fluid tracers. For particles with $St=0.20$, $0.83$, and $9.14$, hyperballistic motion ($\left\langle r^2(t)\right\rangle$ $\propto t^\mu$ with $\mu>2$) is observed at short times, indicating flow-induced acceleration where displacement follows ${\bf r}(t) - {\bf r}(0) =  \frac{1}{2}{\bf a}t^2$ with ${\bf a}$ representing the flow-induced acceleration. More significantly, these particles exhibit superdiffusive behavior ($\left\langle r^2(t)\right\rangle$  $\propto t^\mu$ with $1<\mu<2$) when $t > 10$ rather than converging to normal diffusion. Table \ref{tab:exponents} shows the transport exponent $\mu$ for different Stokes numbers when $t > 10$.  
    
   Superdiffusive transport in turbulent flows is often attributed to a combination of trapping in coherent structures and long-distance flight events. To quantitatively analyze particle trapping in vortical structures, we employ the Okubo--Weiss criterion, based on the decomposition of the fluid velocity gradient~\cite{Weiss1991}:
    \begin{equation}
    Q=s^2-\omega^2,
    \end{equation}
    where $s^2=s_1^2+s_2^2$ with,
    $s_1=\partial_x u_x-\partial_y u_y, s_2=\partial_x u_y+\partial_y u_x$ and $ \omega=\partial_x u_y-\partial_y u_x$.
    Using a threshold $Q_0=\sqrt{\langle Q^2\rangle}$, the flow domain is partitioned into three regions: strongly elliptic regions where $Q \leq -Q_0$ (vorticity dominated), strongly hyperbolic regions where $Q \geq Q_0$ (deformation dominated), and intermediate regions where $-Q_0 < Q < Q_0$.
    
\begin{figure}
      \includegraphics[width=0.97\columnwidth]{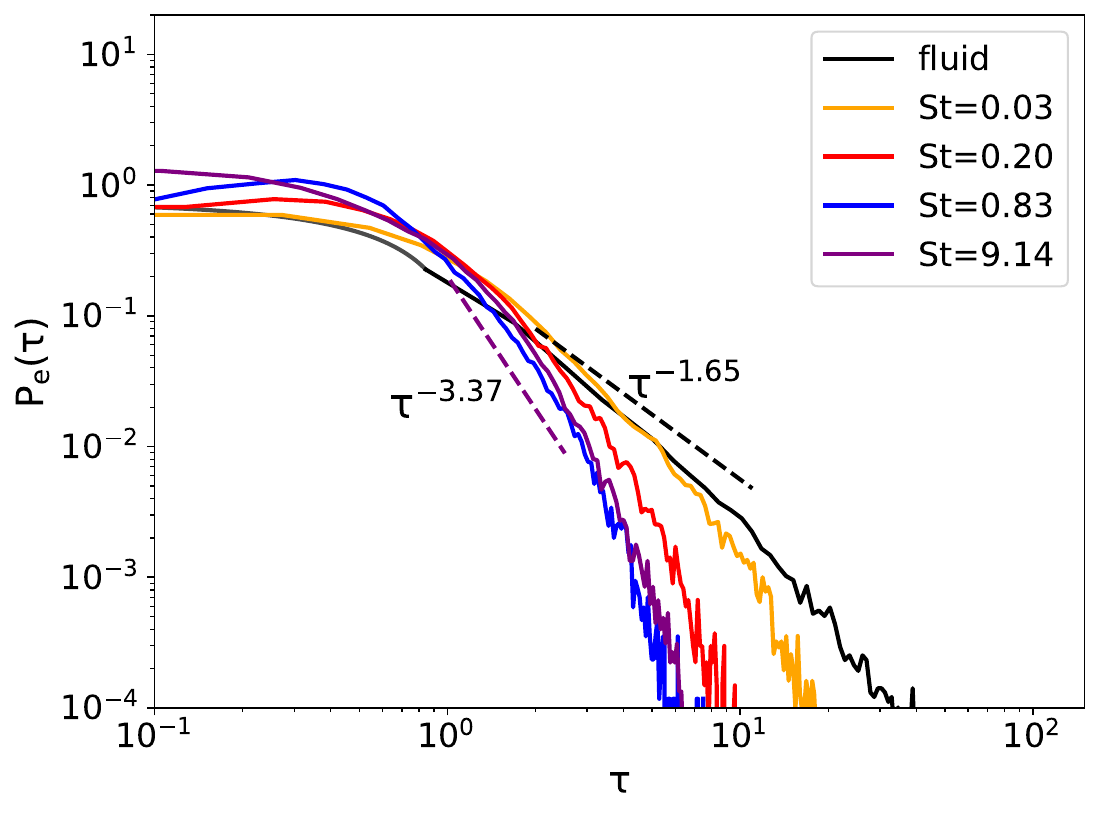}
      \caption{\label{fig:residence_time}PDFs of trapping times in strongly elliptic regions (vortices) for particles with different Stokes numbers. Fluid tracers ($St=0$) show longer trapping times compared to particles with non-zero Stokes numbers, indicating that inertial effects enable particles to escape vortical trapping more easily.}
    \end{figure}

    Figure~\ref{fig:residence_time} shows probability density functions (PDFs) of trapping times in strongly elliptic regions (vortices) for different Stokes numbers. Fluid tracers exhibit longer residence times, while all inertial cases show a steeper decay, indicating that inertia enables particles to escape vortices more easily. Figure~\ref{fig:residence_time} also illustrates a power-law scaling characterized by the exponent $\gamma$, with the values assembled in Table~\ref{tab:exponents}.
 
   \begin{figure}
      \includegraphics[width=0.8\columnwidth]{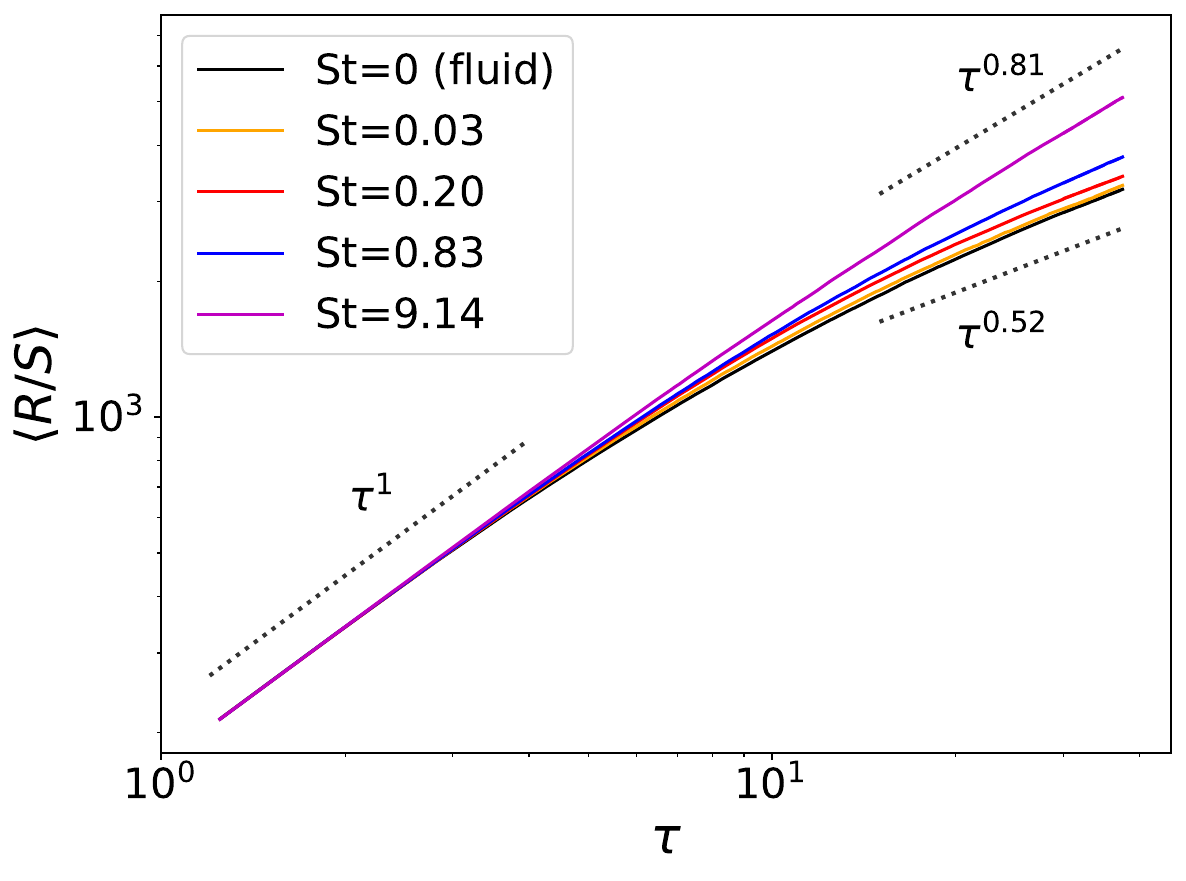}
    \caption{\label{fig:Hurst exponent} Log-log plot of the Rescaled Range (R/S) versus time lag ($\tau$) for different Stokes numbers.}
    \end{figure}
 
 The connection between trapping time statistics and the superdiffusion is a key feature of the fractional kinetics framework, which describes anomalous transport in chaotic systems \cite{Zaslavsky1994, Zaslavsky1997}. This theory predicts a direct relationship between the exponent of the power-law trapping time distribution $\gamma$, and the MSD exponent $\mu$, given by  $|\gamma| = \mu + 2$ \cite{Zaslavsky1997}. Our simulations provide a strong test of this relationship. For the cases identified as superdiffusive, we find  agreement with the theoretical prediction within numerical errors. Specifically, for $St=0.83$, we find $\mu \approx 1.32$ and $|\gamma| \approx 3.37$ (vs. predicted $3.32$), and for $St=9.14$, we find $\mu \approx 1.21$ and $|\gamma| \approx 3.19$ (vs. predicted $3.21$).   This does not hold for near-diffusive cases ($St \le 0.20$). Let us note that the observation that the mean trapping time decreases with increasing Stokes number, does not prove superdiffusion, since it could also be consistent with a larger diffusion coefficient in a normal  diffusion process. However, the successful validation of the  $|\gamma| = \mu + 2$ relationship is a powerful consistency check. This relationship is a hallmark of transport governed by fractional kinetics and is not expected in a simple diffusive process. As anticipated the relationship holds less well for the near-diffusive cases ($St \le 0.20$) as shown in Table~\ref{tab:exponents}.

        To better understand the superdiffusive behavior, we performed a Hurst analysis on the Lagrangian velocity time series. The standard Hurst exponent  has been used for years to quantify correlations. This method is known for its strong resilience to random noise \cite{Carreras1998, Mier2008}. There are many ways to compute the Hurst exponent. One of them is the R/S method \cite{hurst1951}. The R/S method consists of constructing the rescaled range of a time series, $\{V_k, k=1, \dots, N\}$, which is defined as:
\begin{equation}
[\mathrm{R} / \mathrm{S}](\tau)=\frac{\max _{1 \leqslant k \leqslant \tau} W(k, \tau)-\min _{1 \leqslant k \leqslant \tau} W(k, \tau)}{\left(\left\langle (V - \langle V\rangle_\tau)^2 \right\rangle_\tau\right)^{1 / 2}}.
\end{equation}
Here, $W(k, \tau)=\sum_{i=1}^k\left(V_i-\langle V\rangle_\tau\right)$ is the cumulative sum of the demeaned series, and the denominator is the standard deviation over the time lag $\tau$. The term $\langle V\rangle_\tau$ represents the average   of the series $V_i$ over the time lag $\tau$. If the signal is self-similar, the rescaled range follows a power law, $[\mathrm{R} / \mathrm{S}](\tau) \sim \tau^H$, which allows the Hurst exponent, $H$, to be determined from the slope of a log-log plot of $[R/S]$ versus $\tau$. An exponent of $H=0.5$ signifies an uncorrelated random process typical of standard diffusion, whereas $H>0.5$ indicates persistence, a positive long-range correlation. Figure~\ref{fig:Hurst exponent} shows the  R/S analysis for the Lagrangian velocity time series, plotted for all Stokes numbers in log-log scale. At short time lags ($\tau < 10$), all curves collapse onto a single line with a slope corresponding to a Hurst exponent of $H \approx 1$. This indicates a process with perfect persistence, where the particle velocity is completely correlated with itself over these short durations.  
In the long-time asymptotic regime ($\tau > 10$), the curves diverge, and their slopes, which represent the long-range Hurst exponent, become dependent on the Stokes number
and are summarized in Table~\ref{tab:exponents}.
They provide conclusive evidence of an inertia effect.  For the fluid tracers ($St=0$), we find $H \approx 0.52$, a value consistent with the nearly random motion expected for a diffusive process. However, as the Stokes number increases, the Hurst exponent likewise increases, reaching $H \approx 0.81$ for $St = 9.14$. This monotonic increase of $H$ provides strong quantitative proof that particle inertia induces strong long-range persistence in the particle trajectories.

  The autocorrelation function of Lagrangian velocities is also analysed. The autocorrelation $R(\tau) = \langle \mathbf{v}_p(t)\cdot \mathbf{v}_p(t+\tau)\rangle / \sigma_{v_p}^2$
     where $\tau$ is the time lag and $\sigma_{v_p}^2$ represents the variance of the particle velocity.   The Taylor microscale, $ \tau_\lambda$, is determined from the curvature of $R(\tau)$ at $\tau=0$, and is defined as : $\tau_\lambda = \left(-\frac{1}{2} R''(0)\right)^{-1/2}.$ The Taylor microscale characterizes the rapid decay of the correlation near $\tau = 0$ and is associated with the small vortices in the flow. Table~\ref{tab:tau_gamma} presents the  values of the Taylor microscale for different Stokes numbers.  We observe that $ \tau_\lambda$ increases as Stokes number increases, meaning  their  velocity correlations decay more slowly near zero lag and  particles less influenced by rapid fluctuations. The high Stokes number act as a low-pass filter, losing the ability to respond to smaller eddies.  This loss of response to small eddies prevents them from being easily trapped and allows them to maintain their momentum for longer, contributing to the persistent motion identified by the Hurst analysis.

    \begin{table}[th!]
    \caption{\label{tab:tau_gamma}Taylor microscale $\tau_\lambda$ for different Stokes numbers.}
    \begin{ruledtabular}
    \begin{tabular}{cccccc}
    $St$       & 0 (fluid) & 0.03   & 0.20   & 0.83   & 9.14 \\[0.5ex]
    \hline \\[-1.5ex]
    $\tau_{\lambda}$  & 0.46   & 0.48 & 0.49 & 0.52 & 1.06 \\
    \end{tabular}
    \end{ruledtabular}
    \label{tab:mu_gamma}
    \end{table}
   Our analysis of heavy impurity transport in drift-wave turbulence reveals distinct transport behaviors depending on Stokes number. Lower-charge tungsten impurities (high Stokes numbers) and intermediate-charge  (intermediate Stokes numbers) exhibit superdiffusive transport, shorter trapping times in vortices and larger Hurst exponents. In contrast, highly charged tungsten impurities (low Stokes numbers) behave similarly to fluid tracers with longer vortex trapping and normal diffusive transport at long time scales. Our results highlight inertial effects as a crucial factor contributing to anomalous transport in fusion plasmas, particularly for heavy impurities like tungsten. These findings have implications for understanding and predicting impurity transport in fusion devices. The enhanced transport of lower-charge and intermediate-charge tungsten impurities suggests that these species may propagate more rapidly through the plasma, accelerating their accumulation in the plasma core and degrade the plasma confinement. These findings are contextualized within the simplifying assumptions of the HW model. The two-dimensional HW framework is employed to isolate and rigorously characterize the fundamental role of particle inertia in anomalous transport. The limitation is that  the model neglects several effects critical for a full, quantitative description of transport in operational fusion devices like ITER, including parallel  dynamics, magnetic drifts and toroidal geometry, etc. Future work will extend this analysis to incorporate these  effects  to refine the application of these findings to fusion reactor scenarios.

 ZL, KS and SB acknowledge the financial support from  the French Federation for Magnetic Fusion Studies (FR-FCM) and the Eurofusion consortium,  funded by the  Euratom  Research and Training Programme under Grant Agreement No. 633053. ZL, BK and KS acknowledge partial funding from the Agence Nationale de la Recherche (ANR), project CM2E, grant ANR-20-CE46-0010-01. Centre de Calcul Intensif d’Aix-Marseille is acknowledged for providing access to its high performance computing resources.
    \nocite{*}
    \bibliography{revised_manuscript}%

@article{Pitts2013,
  author = {R. A. Pitts and others},
  title = {A full tungsten divertor for ITER: Physics issues and design status},
  journal = {J. Nucl. Mater.},
  volume = {438},
  pages = {S48--S56},
  year = {2013},
  doi = {10.1016/j.jnucmat.2013.01.008}
}

@article{Hollmann2015,
  author = {E. M. Hollmann and others},
  title = {Status of research toward the ITER disruption mitigation system},
  journal = {Phys. Plasmas},
  volume = {22},
  number = {2},
  pages = {021802},
  year = {2015},
  doi = {10.1063/1.4901251}
}

@article{Angioni2021,
  author = {C. Angioni},
  title = {Impurity transport in tokamak plasmas, theory, modelling and comparison with experiments},
  journal = {Plasma Phys. Control. Fusion},
  volume = {63},
  number = {7},
  pages = {073001},
  year = {2021},
  doi = {10.1088/1361-6587/abfc9a}
}

@article{Kallenbach2013,
  author = {A. Kallenbach and others},
  title = {Impurity seeding for tokamak power exhaust: from present devices via ITER to DEMO},
  journal = {Plasma Phys. Control. Fusion},
  volume = {55},
  number = {12},
  pages = {124041},
  year = {2013},
  doi = {10.1088/0741-3335/55/12/124041}
}

@article{Naulin2007,
  author = {V. Naulin},
  title = {Turbulent transport and the plasma edge},
  journal = {J. Nucl. Mater.},
  volume = {363},
  pages = {24--31},
  year = {2007},
  doi = {10.1016/j.jnucmat.2007.01.007}
}

@article{Gustavsson2016,
  author = {Gustavsson, B. and others},
  journal = {Adv. Phys.},
  volume = {65},
  pages = {1},
  year = {2016}
}

@article{Futatani2008,
  author = {Futatani, S. and others},
  journal = {Phys. Rev. Lett.},
  volume = {100},
  pages = {025005},
  year = {2008}
}

@article{Lin2025a,
  author = {Z. Lin and others},
  journal = {J. Plasma Phys.},
  year = {2025},
  volume = {91},
  number = {1},
  pages = {E30},
  doi = {10.1017/S0022377824001259}
}

@article{Lin2025b,
  author    = {Z. Lin and others},
  title     = {},
  journal   = {Plasma Phys. Control. Fusion},
  volume    = {67},
  number    = {4},
  pages     = {045038},
  year      = {2025},
  month     = {apr},
  publisher = {IOP Publishing},
  doi       = {10.1088/1361-6587/adc508},
  url       = {https://dx.doi.org/10.1088/1361-6587/adc508}
}

@article{Krieger1999,
  author = {Krieger, K. and others},
  journal = {J. Nucl. Mater.},
  volume = {266},
  pages = {207},
  year = {1999}
}

@article{Yao2015,
  author = {Yao, J. and others},
  journal = {Fusion Eng. Des.},
  volume = {98},
  pages = {1692},
  year = {2015}
}

@article{Pitts2009,
  author = {Pitts, R. A. and others},
  journal = {Phys. Scr.},
  volume = {T138},
  pages = {014001},
  year = {2009}
}

@article{Weiss1991,
  author = {Weiss, J.},
  journal = {Physica D},
  volume = {48},
  pages = {273},
  year = {1991}
}

@article{Bos2010,
author = {W. Bos and others},
title = {Lagrangian dynamics of drift-wave turbulence},
journal = {Physica D},
year = {2010},
volume = {239},
pages = {1269–1277}
}

@article{Hasegawa1983,
author = {A. Hasegawa and M. Wakatani},
title = {Plasma edge turbulence},
journal = {Phys. Rev. Lett.},
year = {1983},
volume = {50},
pages = {682--686}
}

@article{horton1999drift,
  author  = {W. Horton},
  journal = {Rev. Mod. Phys.},
  volume  = {71},
  number  = {3},
  pages   = {735},
  year    = {1999}
}

@article{Hidalgo1995,
doi = {10.1088/0741-3335/37/11A/004},
url = {https://dx.doi.org/10.1088/0741-3335/37/11A/004},
year = {1995},
month = {nov},
publisher = {},
volume = {37},
number = {11A},
pages = {A53},
author = {C Hidalgo},
title = {Edge turbulence and anomalous transport in fusion plasmas},
journal = {Plasma Phys. Control. Fusion},}

@article{Zaslavsky1994,
  author = {Zaslavsky, G. M.},
  title = {Fractional kinetic equation for Hamiltonian chaos},
  journal = {Physica D},
  volume = {76},
  number = {1-3},
  pages = {110--122},
  year = {1994},
  doi = {10.1016/0167-2789(94)00044-Q},
  publisher = {Elsevier}
}

@article{Zaslavsky1997,
  author = {G. M. Zaslavsky and others},
  journal = {Chaos},
  volume = {7},
  pages = {159},
  year = {1997},
  doi = {10.1063/1.166252}
}

@article{Kadoch2010,
  author = {B. Kadoch and others},
  journal = {Phys. Rev. Lett.},
  volume = {105},
  pages = {145001},
  year = {2010},
}

@article{Sanchez2008,
  author = {Sánchez, R. and others},
  journal = {Phys. Rev. Lett.},
  volume = {101},
  pages = {205002},
  year = {2008},
}

@article{camargo1995,
  title={Resistive drift-wave turbulence},
  author={Camargo, S J and others},
  journal={Phys. Plasmas},
  volume={2},
  number={1},
  pages={48--62},
  year={1995},
  publisher={American Institute of Physics}
}

@article{Lin2024,
  title={Synthesizing impurity clustering in the edge plasma of tokamaks using neural networks},
  author={Z Lin and others},
  journal={Phys. Plasmas},
  volume={31},
  number={3},
  year={2024},
  publisher={AIP Publishing}
}

@article{Carreras1998,
  author = {Carreras, B. A. and  others},
  title = {{Self-similarity of the plasma edge fluctuations}},
  journal = {Phys. Plasmas},
  volume = {5},
  number = {10},
  pages = {3632--3639},
  year = {1998},
  doi = {10.1063/1.873081}
}

@article{Sanchez2015,
  author = {Sanchez, R. and Newman, D. E.},
  title = {{Self-organized criticality and the dynamics of near-marginal turbulent transport in magnetically confined fusion plasmas}},
  journal = {Plasma Phys. Control. Fusion},
  volume = {57},
  number = {12},
  pages = {123002},
  year = {2015},
  doi = {10.1088/0741-3335/57/12/123002}
}

@article{Mier2008,
  title = {Characterization of Nondiffusive Transport in Plasma Turbulence via a Novel Lagrangian Method},
  author = {Mier, J. A. and others},
  journal = {Phys. Rev. Lett.},
  volume = {101},
  issue = {16},
  pages = {165001},
  numpages = {4},
  year = {2008},
  month = {Oct},
  publisher = {American Physical Society},
  doi = {10.1103/PhysRevLett.101.165001},
  url = {https://link.aps.org/doi/10.1103/PhysRevLett.101.165001}
}

@article{Sanchez2006,
  title = {Renormalization of tracer turbulence leading to fractional differential equations},
  author = {S\'anchez, R. and others},
  journal = {Phys. Rev. E},
  volume = {74},
  issue = {1},
  pages = {016305},
  numpages = {11},
  year = {2006},
  month = {Jul},
  publisher = {American Physical Society},
  doi = {10.1103/PhysRevE.74.016305},
  url = {https://link.aps.org/doi/10.1103/PhysRevE.74.016305}
}

@article{hurst1951,
  title={Long-term storage capacity of reservoirs},
  author={Hurst, H. E.},
  journal={Trans. Am. Soc. Civ. Eng.},
  volume={116},
  number={1},
  pages={770--799},
  year={1951},
  publisher={American Society of C}
}
    
    \end{document}